\documentclass[varenna]{cimento}
\title{A BRIEF HISTORY OF OUR UNDERSTANDING OF BEC: FROM BOSE TO 
BELIAEV}
\author{A. Griffin}
\institute{Department of Physics, University of Toronto, Toronto, 
Ontario M5S 1A7, Canada} 
\PACSes{\PACSit{01.65.+g, 03.75.Fi, 05.30.Jp, 67.40.-w}{}}

\begin{document}
\maketitle
\begin{abstract}
We review how our current ideas about BEC developed in the early 
period 1925-1965, which had the specific goal of understanding 
superfluid $^4$He.  This history is presented by commenting on the 
key contributions made by Einstein, Fritz London, Tisza, Landau, 
Bogoliubov, Oliver Penrose and Feynman.  We emphasize the emergence 
of the concept of a macroscopic wavefunction describing the 
condensate.  Starting with the fundamental work of Beliaev in 1957, 
the period 1957-1965 was a golden era for theoretical studies of 
interacting Bose-condensed gases.  This work provided a sound 
conceptual basis for understanding the properties of trapped atomic 
gases which were discovered thirty years later.
\end{abstract}

\section{Introduction}

I was asked to give an opening lecture about the history of BEC, the 
subject of this Varenna Summer School.  The complete history of BEC 
is of general interest and someone should write a popular book with 
the title:   {\it The Quest for BEC:  from Bose to Boulder 
(1925-1995)}.  Today I will give you a truncated version of this 
history, covering mainly the early period 1925 - 1965, from the point 
of view of a condensed matter theorist.
This type of lecture is always scary to give since it involves  
value-judgements about the significance of the work of others.  
However, I think there are several good reasons to review some of the 
highlights of the early search for BEC and how we think about it:

\begin{enumerate}
\item This lecture is like a musical overture, setting out some 
themes (and conflicts) which will be repeated in the rest of this 
book.

\item The story is an exciting one, and many of the players were 
among the greatest physicists of the 20th century.   Moreover, many 
of the key concepts of modern physics (elementary excitations, 
collective modes, broken symmetry, order parameter, etc) were first 
introduced  in dealing with superfluid $^4$He.

\item The history is a complicated one for newcomers to follow 
(especially for those coming from atomic and laser physics) and 
perhaps some guidance is needed.  It involves great theorists like 
Landau and London who had completely different visions about what was 
going on in superfluid $^4$He.  Current work on trapped Bose gases 
has analogous conflicting philosophies.  In particular, the 
literature on superfluid $^4$He has three major rivers which often 
have little overlap:

\begin{itemize}

\item BEC and a macroscopic wave function : London $\rightarrow$
\item Many-body wavefunctions  $\Psi (r_1, r_2,\dots r_N)$ for N 
atoms:  Feynman $\rightarrow$
\item Phenomenological two-fluid theory based on conservation laws 
and a quasiparticle description: Landau $\rightarrow$
\end{itemize}

When we understand these differences, it is easier to understand the 
older literature.  Moreover, if you don't appreciate this history, 
you may ``repeat it'' in trapped gases!

\item A major theme of this lecture will be the emergence of the idea 
of a macroscopic wavefunction $\Phi({\bf r}, t)$ to describe the 
unique features which arise in Bose-condensed fluids. The idea 
started with London in 1938 and by 1965, theorists had developed a 
fairly complete ``conceptual'' understanding of Bose-condensed fluids 
in terms of  a broken-symmetry order parameter.  This required the 
development of field-theoretic techniques in the late 1950's.  The 
period 1957-1965 was a ``golden period" of the theory of Bose liquids 
and gases.  It is this literature and the associated ideas that have 
allowed us to make rapid progress on the theory of trapped atomic 
gases since their dramatic discovery in 1995.
\end{enumerate}

I hope that this sketchy and very personal history of one of the most 
exciting developments in 20$^{th}$ century physics will encourage 
professional historians of science to do a more thorough job in the 
future.  The lecture is presented mainly in the form of separated 
comments on some aspects, {\it to emphasize that it does not claim to 
be a systemmatic study}.  As for some general references where you 
can obtain more detailed information, I would like to suggest:

\renewcommand{\labelenumi}{(\alph{enumi})}
\begin{enumerate}
\item Chapter 1 of my book \cite{ref:griffin} on the modern theory of 
excitations in interacting Bose-condensed fluids, which includes 
extensive references.

\item The recent scientific biography of Fritz London by Gavroglu 
\cite{ref:gavroglu}, who had access to all of London's correspondence 
and private papers.

\item The engaging autobiography of Andronikashvilii \cite{ref:Andron 
89}, who gives an insider's day-by-day account of the low temperature 
group under Landau and Kapitza in Moscow in the period 1935-1960 . 

\item The Ph.D. thesis of Jurkowitz \cite{ref:jurkowitz}, which 
examines how the modern theories of superfluidity and 
superconductivity developed in the 1930's, in the larger context of 
the development of quantum mechanics.

\item The collection of papers by P.W. Anderson \cite{ref:anderson1}, 
with comments by the author. Anderson probably deserves the most 
credit for his pioneering work (in the period 1957-1963) which 
clarified the fundamental role of a macroscopic wavefunction in the 
theory of superfluid $^4$He as well as in superconductors.  I 
particularly recommend the reprinted articles on p.144 and p.166 in 
Ref.\cite{ref:anderson1}.
\end{enumerate}
\renewcommand{\labelenumi}{\arabic{enumi}.}

\section{A. Einstein}

\begin{itemize}
\item Einstein's famous paper \cite{ref:einstein1} was built on a 
paper by S.N. Bose in 1924,  which gave a novel derivation of photon 
statistics and the Planck distribution.  BEC was specifically 
discussed in the second of three great papers by Einstein on the 
statistical mechanics of an ideal monatomic gas (written in 
1924-25).  Pais gives a detailed account of these papers (see p.197 
of Ref.\cite{ref:pais}).

\item Einstein's paper appeared about a year before the development 
of quantum mechanics.  It was the first time anyone referred to or 
used de Broglie's new idea of {\it matter waves}.  Einstein justified 
applying Bose's calculation using the argument that if particles were 
waves, they should obey the same statistics as photons.   Schrodinger 
first heard about de Broglie's idea from reading Einstein's paper.  
So we might say that Schrodinger's wave equation grew out of 
Einstein's  BEC paper, and not the other way around!

\item Einstein's work also preceded the concept of Fermi statistics 
\cite{ref:fermi}, as well as the division of all particles into two 
classes (Fermions and Bosons) depending on their net spin (this was 
cleared up by 1927).

\item The 1925 paper was written when Einstein was 46.  As far as I 
know, Einstein never publically commented about this paper or topic 
again. In the 1930's, Einstein became increasingly interested in the 
interpretation and completeness of quantum mechanics 
\cite{ref:einstein2}.  In recent years, there has been a resurgence 
of research on these questions.  In this connection, the macroscopic 
wavefunction associated with a Bose-condensed gas may allow direct 
tests of certain aspects of quantum mechanics.  So we have come 
around full circle.

\item While the properties of normal Bose gases were extensively 
studied in the decade after Einstein's 1925 paper, nothing much 
happened concerning BEC until 1938.  Apparently a major reason was 
because  George Uhlenbeck (a graduate student of Paul Ehrenfest) 
criticized Einstein's prediction of a ``phase transition'' by arguing 
it would not occur in a finite system (this may remind you of similar 
criticisms in the last few years in trapped Bose gases!).  
Second-order phase transitions were not understood yet and 
Uhlenbeck's criticism was generally accepted (apparently even by 
Einstein, according to the comments by Uhlenbeck - see p.524 of 
Ref.\cite{ref:pais}).  I should add that 
Uhlenbeck introduced the concept of spin and also became a leading 
expert in the kinetic theory of interacting gases and statistical 
mechanics.  In particular, Uhlenbeck and his student Uehling gave a 
very detailed study of the transport coefficients of Fermi and Bose 
gases as early as 1933 \cite{ref:uehluhlen}.
\end{itemize}

\section{Fritz London}
\begin{itemize}
\item London began his University studies studying philosophy then 
switched to theoretical physics as a graduate student.  Is there a 
message here?

\item In the period 1935-37, London (and his brother Heinz London) 
introduced a new theory of superconductivity based on the idea of a 
``macroscopic wave-function"  These ideas underlie and strongly 
influenced the development of the modern microscopic theory of BCS, 
as Bardeen has emphasized many times (in particular, see the epilogue 
written by Bardeen for Gavroglu's biography of London 
\cite{ref:gavroglu}).

\item At a major statistical mechanics conference held in Amsterdam 
in late 1937, London heard discussions which finally clarified the 
nature of second-order phase transitions (they require the 
thermodynamic limit V$\rightarrow\infty$ to be defined in a rigorous 
way) and also that Uhlenbeck had taken back his criticism of 
Einstein's 1925 work \cite{ref:uhlenkahn}.   This got London 
interested in Einstein's forgotten paper on BEC.

\item At the same time,  London also heard rumors of experimental 
work showing superfluidity in liquid $^4$He, below the transition at 
$T_c \sim 2.17K.$  This transition showed a peak in the specific heat 
but it's physical origin was still a mystery.  Indeed, by the early 
1930's, there was increasing evidence that below $T_c$, Helium II (as 
it had been christened)  behaved in strange ways (sudden absence of 
boiling below $T_c$ \cite{ref:mclensmiwil}, infinite thermal 
conductivity, zero viscosity in small channels, etc).   London knew 
$^4$He was a Boson $(S = 0)$ and immediately put the two ideas 
together: {\it Some sort of BEC was involved in the strange phase 
transition shown by superfluid} $^4$He.  As evidence, London noted 
that Einstein's formula for $T_{\rm BEC}$ gave a good estimate of the 
observed transition temperature and the specific heat of the ideal 
gas had a peak at $T_{\rm BEC}$. 

\item London told  Tisza about this idea and after one restless night 
\cite{ref:gavroglu}, Tisza came up with the  two-fluid concept, 
namely that the {\it Bose condensate acted as a new collective degree 
of freedom,  which could move coherently without friction and hence 
give rise to superfluid behaviour}.

\item  Most of this work  was done while London had a temporary 
research position in Paris at the Institut Henri 
Poincar$\acute{\mbox{e}}$.  He was looking for a permanent job 
outside of Germany, where Hitler's Nazi government was starting to 
show its real intentions. London finally accepted a professorship  in 
the Department of Chemistry at Duke University in the USA in 1939, 
where he remained for the rest of his career.

\item The year 1938 was a busy period.  The above-mentioned 
statistical mechanics 
Conference in Amsterdam was in November,  1937.  In January, 1938, 
Kapitza \cite{ref:kapitza} and, independently, Allen and Misener 
\cite{ref:allen}, published their key experiments on superfluidity in 
the same issue of {\it Nature}.   London had his idea in late 
January, submitted a one page letter to {\it Nature} in early March, 
which was duly published in April 9, 1938 \cite{ref:london1}.  London 
discussed his idea with Laszlo Tisza in late January, who then 
submitted a brief note to {\it Nature} on April 16, which was 
published on May 21, 1938 \cite{ref:tisza1}.  We can all admire the 
short time needed for publication in those days.

\item The first paper by London suggesting the relevance of BEC to 
liquid $^4$He was just a few paragraphs but created considerable 
interest \cite{ref:gavroglu}.  More detailed papers were published 
soon after by both London \cite{ref:london2} and Tisza 
\cite{ref:tisza2}. London was never able to develop his  idea about 
BEC in liquid $^4$He in a quantitative manner, and was tremendously 
frustrated by this over the next decade.  He had a vision of the 
correct theory but the techniques needed to fill in the details had 
to wait until the late 1950's, when many-body theory was developed.  
However, modern microscopic theories of superfluidity and 
superconductivity basically vindicate London's concepts and 
philosophy \cite{ref:griffin, ref:anderson1}.  In a strange way, his 
lack of a microscopic model was an advantage since it forced London 
to think deeply about the general features a successful theory of 
superfluidity should have and how these would be mirrored in a 
``macroscopic'' wavefunction.  From his point of view, BEC was 
important because it illustrated how one could produce macroscopic 
coherence effects directly related to quantum mechanics. 

\item Earlier in his career, London worked with Schrodinger as a 
research assistant and shared the latter's idea that the wavefunction 
in quantum mechanics represented something ``real''.  London's 
thinking  about a macroscopic wavefunction that would describe 
superfluids and superconductors was clearly influenced by this.  
However, this approach was not consistent with the developing 
paradigm of the 1930's.  The emphasis on operators in quantum 
mechanics and the resulting de-emphasis in the Copenhagen 
interpretation on ``pictures"  made the tentative ideas of London and 
Tisza seem very old-fashioned (evidence for this view is discussed  
in the interesting Ph.D. thesis of Jurkowitz \cite{ref:jurkowitz}). 
The theoretical ideas of London really only got ``moving''  with the 
later post-war work of Bogoliubov in 1947 \cite{ref:bogoliubov1} and 
Oliver Penrose in 1951 \cite{ref:penrose}, and  finally bought to 
completion in the early 1960's.
\end{itemize}

\section{L. Tisza}
\begin{itemize}
\item Lazlo Tisza worked in Landau's group in Kharkov during 
1935-1937.  In the period 1937-1938, he was a visiting research 
fellow at the Coll\`ege de France in Paris, where he interacted with 
London.  Tisza eventually went to MIT in the U.S.A., where he has 
spent the rest of his career.

\item Tisza had nothing but an ideal Bose gas as a microscopic model, 
yet he developed a ``two-fluid hydrodynamics" based on the notion of 
a superfluid and normal fluid.  
Tisza's two long papers published in 1940 on his two-fluid 
hydrodynamics are very impressive even today \cite{ref:tisza2}.  He 
could explain all the experiments exhibiting superfluidity, usually 
involving the normal fluid and superfluid moving in opposite 
directions.  Tisza also predicted the existence of a new kind of 
hydrodynamic oscillation,  a temperature wave (later called second 
sound by Landau).

\item London took many years \cite{ref:gavroglu} to accept the huge 
``leap'' that Tisza made connecting his BEC idea with superfluid 
behaviour based on a two-fluid model.  Personally, I have been always 
puzzled by this reluctance.  In any event, they remained close 
friends and corresponded at length \cite{ref:gavroglu} on BEC-related 
questions until London died in 1954.

\end{itemize}

\section{L.D. Landau}

\begin {itemize}
\item Landau's ``bombshell'' of a paper \cite{ref:landau1} on 
superfluid $^4$He changed how we think about {\it all} condensed 
matter systems.

\item While essentially phenomenological, Landau introduced a ``new'' 
hydrodynamics to describe low frequency superfluid phenomena based on 
the motion of two fluids described by $\rho_n$, $\rho_s$ ${\bf v}_n$, 
${\bf v}_s$,  in many ways similar to those developed earlier by 
Tisza \cite{ref:tisza2}.

\item In addition, Landau also introduced the novel but powerful idea 
that the liquid could be described in terms of a ``gas of weakly 
interacting quasiparticles", with a relatively simple energy spectrum 
for two kinds of quasiparticles: phonons and rotons.  This 
quasiparticle idea allowed Landau to do quantitative calculations.  
Landau orginally tried to justify his quasiparticle energy spectrum 
using ``quantum hydrodynamics" but this was never convincing.  In a 
later brief addendum \cite{ref:landau2}, Landau introduced the now 
famous phonon-roton spectrum as a single excitation branch.  This 
modified spectrum was deduced using fits to better thermodynamic data 
which had become available.  

\item Landau also introduced the idea of collective modes as distinct 
from quasiparticles (or elementary excitations).  In particular, 
first and second sound are collective modes in the ``gas of 
quasiparticles".

\item By the early 1950's, the Landau-Khalatnikov (LK) theory had 
overshadowed the London-Tisza scenario,  which still lacked a 
microscopic model in which interactions between atoms were included. 
The direct measurement of the phonon-roton spectum using inelastic 
neutron scattering \cite{ref:griffin, ref:yarnarnbendkerr, 
ref:henwoods} dramatically confirmed the correctness of the Landau 
approach.  To this day, the Landau-Khalatnikov  theory of superfluid 
$^4$He (as summarized in the classic book by Khalatnikov 
\cite{ref:khalatnikov}) is the standard theory used to describe the 
properties of superfluid  $^4$He. In his landmark paper, however, 
Landau never mentioned the fact that $^4$He atoms were Bosons, let 
alone the existence of a Bose condensate.   Indeed, the experimental 
low temperature community still tends to feel BEC cannot play a very 
fundamental role since the LK theory  hardly mentions it! On the 
other hand, since the 1960's, most theorists have viewed the LK 
theory as a phenomenological theory whose microscopic basis lies in 
the existence of a  Bose macroscopic order parameter.

\item Why did Landau resist the idea of a Bose condensate as being 
relevant to superfluid $^4$He?  This is strange, since it was Landau 
himself who in 1937 formulated the usefulness of the concept of an 
{\it order parameter} to deal with second-order phase transitions.  
However, even as late as 1949, Landau wrote a blistering attack on 
Tisza's work in the Physical Review \cite{ref:landau3}.  Overlooking 
personality conflicts (see Ref.\cite{ref:gavroglu}), one interesting 
explanation is suggested by Jurkowitz \cite{ref:jurkowitz}.  Landau 
deeply accepted the Copenhagen view of quantum mechanics, as is clear 
from Chapter 1 of Ref.\cite{ref:landau4}. In particular, he accepted 
that the proper procedure was to take a classical description and 
quantize it by introducing operators for the physical observables.  
Thus, Landau believed that the correct way of understanding a 
``quantum" liquid was to {\it quantize} the standard hydrodynamical 
theory of a ``classical" liquid.  This is what he tried to do with 
his ``quantum hydrodynamics" in his 1941 paper - an approach that, as 
mentioned above,  never succeeded. However, it suggests why  in the 
1940's, Landau felt that it was absolutely wrong to try and develop a 
theory of a ``quantum liquid'' starting from a ``quantum gas'' (which 
is what the London-Tisza program was trying to do).  In modern 
renormalization group (R.G.) language, one might say that Landau felt 
these two systems corresponded to two different fixed points.

\item More generally, we know that Landau was deeply imbued with the 
idea that a good theory should be able to give a quantitative 
explanation of experimental data.  He did not like ``vague'' theories 
which could not be pushed to a clear experimental conclusion.  In any 
event, as far as I know, Landau never ``officially" changed his views 
of the London program that a theory of superfluid $^4$He could be 
developed ``starting'' from an ideal Bose gas.  However, towards the 
end of the second of two great papers by Beliaev in 1957, Beliaev 
writes \cite{ref:beliaev1} ``It allows one to suppose that the 
difference between liquid He and a non-ideal Bose gas is only a 
quantitative one, and that no qualitatively new phenomena arise in 
the transition from gas to liquid".  A few paragraphs later, Beliaev 
ends his paper by thanking ``L.D. Landau for criticism of the 
results",  as much of a ``  stamp of approval" as one could expect.
\end{itemize}

\section{N.N. Bogoliubov}

\begin{itemize}
\item The famous paper by Bogoliubov in 1947 \cite{ref:bogoliubov1} 
was a real breakthrough, but it took over a decade  to be generally 
understood or accepted by the condensed matter community.

\item Bogoliubov's calculation showed how BEC was {\it not} much 
altered by interactions in a weakly interacting dilute Bose gas 
(WIDBG), something which was not obvious at the time. 

\item However, Bogoliubov showed that interactions completely altered 
the long wavelength response of a Bose-condensed gas.  The predicted 
phonon spectrum at low momentum was exactly what was {\it assumed} by 
Landau for his quasiparticle dispersion relation and ensured the 
stability of superfluid flow.  

\item This paper took up the sputtering program of London and Tisza, 
and started the developments which led to a ``complete" theory by the 
early 1960's based on the key  role of Bose broken symmetry, and how 
this modified the dynamics of a Bose-condensed system, be it a  gas 
or liquid (for a review with references, see 
Ref.\cite{ref:griffin}).  It seems that London never heard of 
Bogoliubov's work (there is no mention of it in 
Ref.\cite{ref:gavroglu} or in London's well known monograph). 

\item Landau clearly recognized the correctness and importance of 
Bogoliubov's results (see Ref.\cite{ref:landau3}). However, the paper 
of Bogoliubov was not ``understood'' until 1957 or so. This was 
probably mainly due to the use of second quantization (unfamiliar at 
that time in condensed matter physics)  and the use of a number 
non-conserving approximation.  For example, the famous papers of 
Feynman \cite{ref:feynman} as well as those by Lee, Yang and Huang 
(their work is nicely summarized in the {\it first} edition of 
Huang's well known book \cite{ref:huang}) in the middle 1950's make 
essentially no reference to Bogoliubov's paper. The appearance of the 
BCS theory of superconductivity in 1957 \cite{ref:bardcoopscrief}, 
however, quickly led to analogies with Bogoliubov's treatment of a 
dilute Bose gas \cite{ref:anderson1}.  Both theories involve 
``off-diagonal'' symmetry-breaking mean-fields.
\end{itemize}

\section{O. Penrose and L. Onsager}
In liquid $^4$He, it is easy to get into the collision - dominated 
hydrodynamics domain, described by dynamic local thermal 
equilibrium.  Thus all the original experiments in the 1930's were on 
low frequency, hydrodynamic phenomena, as were the theories of Tisza 
and later by Landau. It was the two-fluid domain that was most easily 
studied, a domain described by the normal fluid and superfluid 
variables:  $\rho_n$, $\rho_s$ ${\bf v}_n$, ${\bf v}_s.$   An 
important event was in 1946, when Peshkov in Moscow 
\cite{ref:peshkov} succeeded in observing second sound as a 
temperature oscillation and showed that the temperature dependence of 
the second sound velocity agreed with the prediction of two-fluid 
hydrodynamics.
Direct evidence for BEC (in contrast to superfluidity) was  illusive 
in superfluid $^4$He.  In contrast, in atomic gases,  BEC was 
immediately observed in the first successful experiment 
\cite{ref:anderson2},  while the two-fluid hydrodynamic region is 
difficult to access because of the low density.  Only recently have 
experiments on  trapped gases started to probe this interesting 
two-fluid region \cite{ref:stamper}.

In modern theory, the underlying Bose complex order parameter is 
\[
 \Phi({\bf r}, t) = \sqrt{n_c({\bf r}, t)}\ e^{i\theta{(\bf r}, t)}.
\]
The superfluid velocity, defined by $m{\bf v}_s({\bf r}, t) = 
\hbar\nabla\theta({\bf r}, t), $ is easily measured by a variety of 
experiments.  In contrast, the amplitude $\sqrt{n_c}$ is more 
illusive since it doesn't appear in most measurable properties of 
superfluid $^4$He.  What is easily studied is the superfluid density 
$\rho_s,$ which depends  on $n_c$ but in a complicated  manner.

In 1956, Penrose and Onsager \cite{ref:penonsager} extended the 
concept of $\Phi(\bf r)$ to a uniform Bose {\it liquid} and discussed 
the associated long-range correlations it implied, building on 
earlier work by Penrose \cite{ref:penrose}. They also estimated the 
value of $n_c$ at $T = 0$ using a ground state wavefunction due to 
Feynman and found $n_c\simeq 0.08 n$ in liquid $^4$He.  This estimate 
has not changed much in 40 years!  Convincing experimental values of 
$n_c$ were only obtained in early 1980's, using inelastic neutron 
scattering to extract the momentum distribution  $n_{\bf p}$ of the 
$^4$He atoms (for reviews, see Ch.4 of Ref.\cite{ref:griffin} and the 
article by Sokol in Ref.\cite{ref:grifsmokstring}). These experiments 
also gave (when carefully analyzed) $n_c\sim 0.1n$ at $T=0$ and show 
that $n_c \rightarrow 0$ as $T\rightarrow T_c.$

\section{R.P. Feynman}

\begin{itemize}
\item Several brilliant papers by Feynman \cite{ref:feynman} posed 
the important question: How are the many-body wavefunctions $\Psi_N 
(r{_1}, r{_2},\dots r{_N})$ for the ground state and lowest excited 
states of liquid $^4$He effected by Bose statistics?

\item These papers are essentially variational in nature, but Feynman 
managed to give the first ``microscopic" understanding based on 
quantum mechanics of  the roton part of the quasiparticle spectrum 
(as first postulated without any explanation by Landau in 1947 
\cite{ref:landau2}).

\item Feynman's papers were the beginning of  a huge literature on 
various approximations for the wavefunctions $\Psi_N$ of 
many-particle quantum states. In particular, one should mention the 
extensive work of Feenberg and his students and co-workers in 
developing what is called the 
``correlated basis function'' approach (for a brief review, see 
Section 9.1 of Ref.\cite{ref:griffin}).  This has been mainly 
successful in calculating ground-state properties $(T=0)$. However, 
the role of a Bose condensate in all these theories is obscure. It 
appears, at best, as a sort of ``side effect". Little contact is made 
with the field-theoretic approach based on a macroscopic 
wave-function.  Moreover, by concentrating from the beginning on 
realistic treatment of the hard-core and high density which arise in 
a liquid, these theories do not appear (so far) to have much 
relevance to trapped dilute Bose gases.

\item Feynman also did fundamental work on vortices and their role in 
superfluid $^4$He.
In this connection,  Feynman's work inadvertently popularized the 
idea that somehow ``rotons" were fundamentally connected to 
superfluidity - and moreover, these were  related to vortices.  This 
is still a lingering belief in parts of the low temperature 
community,  with not much justification!  Needless to say, there are 
vortices in dilute Bose gases but no ``rotons''. 
\end{itemize}

\section{Golden era}

\begin{itemize}
\item The ``Golden Period" (1957 - 1965) ends our  history of the 
first phase of the quest for BEC.  In this period, many important  
theorists attacked the interacting Bose-condensed gas problem. It was 
a hot topic during this period and, in my opinion, the final 
theoretical edifice is one of the great success stories in 
theoretical physics \cite{ref:griffin, ref:beliaev1, ref:beliaev2, 
ref:hugenholtz, ref:gavoret, ref:hohenmartin, ref:bogoliubov2}.  
However, until recently it was largely unknown since it was somewhat 
uncoupled from experiments on liquid $^4$He and involved a very 
complex formalism.  

\item Out of these efforts came a way of ``isolating" the profound 
role of a condensate on the response functions of a Bose-condensed 
fluid.  This was an amazing accomplishment since it was developed to 
deal with a Bose liquid. The problems of a liquid had to 
``separated'' off  from the effects of Bose condensation.  These 
studies often used a weakly interacting dilute Bose gas (WIDBG) as an 
illustration.  Thus a huge amount of theoretical insight was gained 
and a whole literature developed about a ``fictitious" system, namely 
a Bose-condensed gas!  Today, more than thirty years later, all this 
``useless'' work is very relevant to trapped atomic gases, apart from 
the need to add a trapping potential.  In another article in this 
book, I give an introductory review of some of this work 
\cite{ref:griffin2}.

\item I should mention the important and beautiful work of Lee, Yang 
and Huang in period 1957-59  on the low temperature properties of a 
dilute hard-sphere Bose gas \cite{ref:huang}.  However, their work 
involved very complex many-body  calculations using exact 
wavefunctions but never dealt with the underlying Bose broken 
symmetry as Bogoliubov and Beliaev did. The final results are correct 
but the method of calculation does not give much insight into the 
physics involved. 

\item A key development relevant to trapped gases was due to 
Pitaevskii in the period 1959-61.  Many of us now are very familiar 
with the famous Gross-Pitaevskii  equation for  $\Phi({\bf r},t)$ 
\cite{ref:pitaevskii}.  However, Pitaevskii's real contribution was 
not this particular equation - rather it lies more in introducing the 
whole idea of a {\it macroscopic wavefunction} which could depend on 
both position and time. Pitaevskii's work, of course, was no doubt 
inspired by the use of a similar ``wavefunction'' by Ginzburg and 
Landau in their pioneering theory of spatially inhomogeneous 
superconductors \cite{ref:ginzlandau}.
\end{itemize}

\section{Concluding remarks}

A history of scientific ideas often tends to give a misleading 
impression of a steady march towards the ``correct'' theory, ignoring 
the doubts even the original thinkers had about their own ideas.  A 
good example concerns the key role that Bose statistics (and the 
resulting condensate) plays in the phenomenon of superfluidity of 
liquid $^4$He.  It was only in 1949 that the first experiments on 
liquid $^3$He (a Fermi liquid) were reported, showing no evidence for 
superfluidity down to 1K.  It is interesting to read the exchange of 
letters \cite{ref:gavroglu} between London and Tisza about these 
results and catch their excitement.  Clearly, even in 1949, they 
still had lingering doubts about their key idea, which were finally 
put to rest by the lack of superfluidity in liquid $^3$He.  In 
passing, we might note that these negative results for liquid $^3$He 
were probably a major reason for the renewed emphasis on the role of 
Bose statistics in superfluid $^4$He in the early 1950's 
\cite{ref:penrose, ref:feynman}.

Finally, beginning with London's work  in 1935-38, I would like to 
recall that our modern understanding of superfluidity in Bose fluids 
has gone hand-in-hand with our increasing understanding of 
superconductivity \cite{ref:anderson1}:
\begin{eqnarray*}
\Phi({\bf r})&=&\langle{\hat\psi}({\bf r})\rangle_{sb} \ \ 
\rightarrow \ \  \mbox{Bose order parameter in Bose superfluids}\ 
\cite{ref:beliaev2}\\  
&=& \langle{\hat\psi}({\bf r}){\hat\psi}({\bf r})\rangle_{sb} \ \ 
\rightarrow \ \  \mbox{Cooper pair amplitude in superconductors}\ 
\cite{ref:gorkov}
\end{eqnarray*}
It was not realized or emphasized in the early years, but the BCS 
theory \cite{ref:bardcoopscrief} really involves a ``BEC of Cooper 
pairs" (see, however, the early paper by Gor'kov \cite{ref:gorkov}).  
This became clear in 1980's, when Leggett \cite{ref:leggett}, 
Nozi$\grave{\mbox{e}}$res \cite{ref:nozieres} and others (for a 
review, see the article by Randeria in Ref.\cite{ref:grifsmokstring}) 
pointed out that in the strong coupling limit, the BCS theory reduces 
to a Bogoliubov theory of a dilute Bose gas composed of small, 
non-overlapping Cooper pairs. 
It is exciting that in trapped Fermion gases with an attractive 
interaction, this BCS phenomenon is once again being addressed (see, 
for example, Ref.\cite{ref:stoofhoubiers})  and that we may be able 
in the future to study this BCS $\rightarrow$ BEC transition in 
exquisite detail by changing the value of the scattering length $a$ 
by working close to a Feshbach resonance 
\cite{ref:inouandstenmiestampket}.

We have come to the end of  the {\it first phase} of BEC studies.  
From the 1960's, an experimental search started in earnest for a {\it 
pure form} of BEC, without the complications of dealing with a {\it 
liquid}.  That is, the goal was to find BEC in a {\it gas}!  The 
major systems that have been studied are (for review of work before 
1995, see the review articles in Ref.\cite{ref:grifsmokstring}):
\begin{itemize}
\item Optically-excited excitons in semiconductors
\item Spin-polarized hydrogen atoms
\item Laser-cooled alkali atoms
\end{itemize}
I leave the writing of this recent history, and especially since 
1995, to someone in the audience!

All of these developments would have made Fritz London, one of the 
most seminal theorists of all time, very happy!  His vision and that 
of Lazlo Tisza has been realized to a degree  they could never have 
predicted sixty years ago.

\acknowledgments
During a thirty year  career working on the theory of superfluid 
$^4$He and superconductivity, I have enjoyed discussing these 
questions with many of my colleagues in condensed matter physics.  
During this Varenna Summer School, I especially appreciated some 
comments by Lev Pitaevskii about the Landau school.  I also 
acknowledge many stimulating discussions with my new colleagues in 
atomic and laser physics, who bring a fresh perspective to the study 
of BEC.  I would be happy to hear from anyone who might have useful 
information to complete or correct the story.  

My research is supported by NSERC of Canada.

\end{document}